\documentclass{aa}
\usepackage{graphicx}
\usepackage{natbib}

\usepackage{txfonts}
\newcommand{\invcm}{cm$^{-1}$}
\newcommand{\kms}{km\,s$^{-1}$}

\newcommand{\sforbidden}{$\lambda10821$ [S\,{\sc i}] }
\newcommand{\sfor}{[S\,{\sc i}] }

\begin{document}
   \title{Sulphur abundances in disk stars as determined from the forbidden $\lambda10821$ [S\,{\sc i}] line\thanks{Based
   on observations obtained at the Gemini Observatory, which is operated by the
AURA, Inc., under a cooperative agreement with the NSF on behalf of
the Gemini partnership: the NSF (US), the PPARC (UK), the NRC
(Canada), CONICYT (Chile), the ARC, CNPq (Brazil), and CONICET
(Argentina).}}
\authorrunning{N. Ryde}
\titlerunning{Sulphur abundances in disk stars}

   \author{Nils Ryde
          }

   \offprints{N. Ryde}

   \institute{Department of Astronomy and Space Physics, Uppsala University, Box 515, SE-751 20 Uppsala,
         Sweden\\
              \email{ryde@astro.uu.se}
             }

   \date{Received ; accepted }


  \abstract
   {}
   {In this paper we aim to study the chemical evolution of sulphur in the galactic disk, using a new optimal abundance
   indicator: the \sfor line at 10821~\AA.
   Similar to the optimal oxygen indicators, the [O\,{\sc i}] lines, the \sfor line has the virtues
   of being less sensitive to the assumed temperatures of the stars investigated and of likely being less
   prone to non-LTE effects than other tracers.}
   {High-resolution, near-infrared spectra of the \sfor line are recorded using the
   Phoenix spectrometer on the Gemini South telescope. The analysis is based on 1D, LTE model atmospheres using a homogeneous set
   of stellar parameters.}
   {The \sforbidden line is suitable for an abundance analysis of disk stars, and the sulphur abundances derived from it are
   consistent with abundances derived from other tracers. We corroborate that, for disk stars, the trend of
   sulphur-to-iron ratios with metallicity is similar to that found for other alpha elements,
   supporting the idea of a common nucleosynthetic origin. }
   {}


   \maketitle
%

\section{Introduction}

To extract information about the nucleosynthesis of the $\alpha$
elements (e.g., Mg, Si, S, and Ca), accurate data on their
abundances for stars in the galactic halo and disk are essential.
Sulphur is one of the least studied $\alpha$ elements, owing to the
paucity of appropriate spectral lines. The diagnostics tools used
for sulphur so far are multiplets of transitions from highly excited
levels, such as the $\lambda\lambda 6743 - 57$, $\lambda\lambda 8694
- 95$, and $\lambda\lambda 9213 - 38$ \AA\ lines (corresponding to
the oxygen triplet at 7775 \AA ). The derived abundances from these
lines are sensitive to the adopted $T_\mathrm{eff}$ of the star and
could be susceptible to non-LTE effects that are difficult to
quantify owing to a lack of adequate atomic data. Thus, these
multiplets may lead to uncertain sulphur abundances. A recent debate
in the literature has questioned the origin of sulphur in the early
universe, as well as whether sulphur follows the evolution of the
other $\alpha$ elements or not, partly owing to these uncertainties,
but also due to the weakness of lines used (see for example
Israelian \& Rebolo 2001; Takada-Hidai et al. 2002; Nissen et al.
2004; Ryde \& Lambert 2004, 2005; Korn \& Ryde
2005)\nocite{israel,takeda,PEN_halo,ryde:04_S,ryde:05_S_poster,korn_ryde:05}.

Sulphur and oxygen belong to the same group in the periodic table
and have the same number of valence shell electrons, which means
that their spectra show some similarities, and similar diagnostic
lines are available for the determination of their elemental
abundances in stars. In the same way as for the $\lambda 6300$ and
$\lambda 6363$ \AA\ [O\,{\sc i }] lines from the ground state, which
have been extensively used and discussed for the determination of
oxygen abundances in stars (see for example Asplund et al. 2004;
Nissen et al. 2002; Garcia P\'erez et al.
2006)\nocite{asplund_oi1,nissen:oi,perez_oi}, the corresponding
\sfor lines should not be susceptible to non-LTE effects, nor should
they be very dependent on the adopted $T_\mathrm{eff}$. These
virtues make it worthwhile to investigate the \sforbidden line as a
spectroscopic tracer of stellar sulphur abundances. The reason why
this line has not previously been used is that it lies beyond the
reach for normal optical CCDs. The near-infrared region also has
several other advantages for an abundance study of cool stars (Ryde
et al. 2005)\nocite{ryde_munchen_review}.

In this \emph{Letter} we present the first attempt to determine
stellar sulphur abundances based on the forbidden sulphur line at
10821.18~\AA\ (in air).  This line is an inter-combination line, an
M1 transition between the triplet ground-state ($^3P_2$) and the
first excited singlet state ($^1D_2$). We explore its use in stars
of the galactic disk since it is important to evaluate this new
preferred diagnostic and to scrutinize the evolution of sulphur in
the disk based on other diagnostics. The abundances of sulphur in
disk stars have earlier been studied by \citet{clegg},
\citet{francois:87},
 \citet{chen}, and \citet{caffau}.

\section{Observations}

The [S\,{\sc i}] line was observed at a high signal-to-noise ratio
($\mathrm{SNR}\approx 200$), in a sample of G and K subgiants and
giants of the galactic disk\footnote{For the purpose of the present
paper we define all stars with metallicity $\mathrm{[Fe/H]}>-1$ as
"disk" stars. The notation $\mathrm{[A/B]}\equiv \log
(N_{\mathrm{A}}/N_{\mathrm{B}})_{\star} - \log
(N_{\mathrm{A}}/N_{\mathrm{B}})_{\mathrm{\odot}}$, where
$N_{\mathrm{A}}$ and $N_{\mathrm{B}}$  are the number abundances of
elements A and B, respectively.}. The homogeneous set of stars was
selected from the library of F5-K7 stars observed with the Elodie
spectrometer (Soubiran et al. 1998)\nocite{elodie_II} and are listed
in Table \ref{stars}. Our high-resolution, near-infrared spectra
were recorded with the Phoenix spectrometer mounted on the Gemini
South telescope. Phoenix is a cryogenic, single order echelle
spectrometer with an Aladdin $512\times 1024$ element InSb detector
array cooled to 35 K (Hinkle et al. 1998)\nocite{phoenix}. We used a
3 pixel wide slit resulting in a spectral resolution of
$R\sim60,000$. The spectra were centered on the \sforbidden line and
covered the 9220-9260\,\invcm\ region. The data were obtained in
service observing mode (Programme ID: GS-2004A-Q-13) over a period
from 5/4 2004 to 23/6 2005.

\begin{figure}
   \centering
  \includegraphics[width=8.5cm]{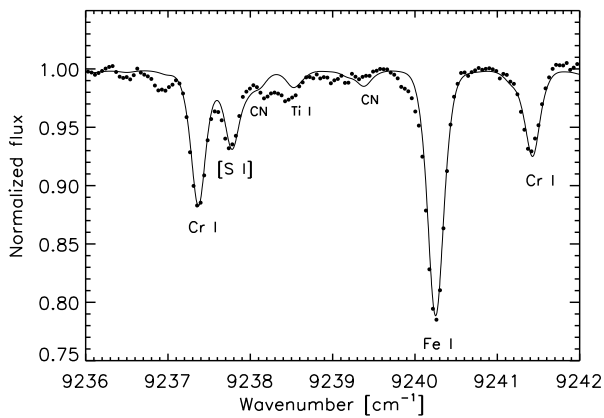}
      \caption{The spectrum of the star HD212943 is shown by dots.  The full line shows a synthetic spectrum
      with 
      $\mathrm{[S/Fe]}=0.21$, based on a model
      atmosphere with the following fundamental parameters: $T_\mathrm {eff}=4586$\,K, $\log g = 2.81$
      (cgs), [Fe/H]$=-0.34$, $\xi_\mathrm{micro}=1.5\,\mathrm{km\,s^{-1}}$, and
      $\xi_\mathrm{macro}=4.5\,\mathrm{km\,s^{-1}}$.
              }
         \label{spectrum}
   \end{figure}

The data were processed in a standard manner with the reduction
package \verb"IRAF" (e.g., Tody 1993)\nocite{IRAF} to retrieve
one-dimensional, continuum-normalized, and wavelength-calibrated
stellar spectra. The raw images were trimmed and a normalized flat
frame was made by dividing a master flat by a Legendre polynomial to
fit the overall spectral shape in the dispersion direction. The
observations were performed in pairs such that the star was moved
approximately $5\arcsec$ along the slit between two equally long
exposures. These two exposures were subsequently subtracted from
each other, a process that reduces the sky background. A hot
calibration star, recording telluric atmospheric absorption lines,
was used partly for the wavelength calibration and partly to
eliminate the telluric spectral contamination. The multiple
exposures of the same star made it possible to apply a
cosmic-ray-hit rejection algorithm when adding them together in the
form of 1-dimensional spectra. Due to residual fringing, a
high-order polynomial was used to rectify and normalize the spectra.
An example of a spectrum, with lines identified, is shown in Fig.
\ref{spectrum}.

In the analysis the equivalent width of the \sforbidden line was
measured using the deblending algorithm in \verb"IRAF", since the
wing of a Cr\,{\sc i} line blends into the wing of the \sfor line.
Gaussian fits were used, and we estimate the measuring uncertainty
to be 10\%, leading to an uncertainty in the derived sulphur
abundance of the order of 0.03 dex, as estimated from individual
measurements. The equivalent widths are given in Table
\ref{spectrum}.

\section{Analysis}

   \begin{table*}
      \caption[]{The stars investigated. Columns 3-6 give the atmospheric parameters of the stars, after \citet{elodie_II}.
      The two subsequent columns provide the dates and the exposure times of the observations of the stars.
      The last four columns give our measured equivalent widths of the \sforbidden line, the derived logarithmic
      number abundance of sulphur relative to hydrogen, the logarithmic sulphur
      abundance relative to the Sun, and finally, the logarithmic sulphur-to-metallicity ratio relative to
      the solar values.
      In the second part of the table the stellar parameters for two stars from \citet{fuhrmann:04}, and
      the sulphur abundances we derive based on these, are presented. We have used a solar sulphur abundance of
      $\log \varepsilon_\odot(\mathrm S)\equiv\log N_\mathrm S/N_\mathrm H + 12=7.20$ \citep{chen}.}
         \label{stars}
   $$
        \begin{array}{llcccccccccc}
            \hline
            \hline
            \noalign{\smallskip}
        \mathrm{Star}    & \mathrm{Spectral}  &  T_{\mathrm{eff}}  & \log(g) & \mathrm{[Fe/H]} & \xi_\mathrm{micro}/\xi_\mathrm{macro} & \mathrm{Date\,\,of} & \mathrm{Exposure\,\, time}& W_\mathrm S & \log \varepsilon(\mathrm S)& \mathrm{[S/H]} & \mathrm{[S/Fe]}\\
            & \mathrm{type}  & \mathrm{[K]}& \mathrm { [cgs]} & &\mathrm{[km\,\, s^{-1}}]   & \mathrm{observations}  & \mathrm{[s}]& \mathrm{[m\AA ]}& &\\
            \hline
            \noalign{\smallskip}
           \textrm {\object{HD~3546}} & \textrm {G5III}  & 4942 & 2.73  & -0.66 & 1.5/ 5.5  &\textrm{July 6, 2004} & 120 & 11.8 & 6.79 & -0.41 & 0.25 \\
          \textrm {\object{HD~10380}} & \textrm {K3III}  & 4057 & 1.43  & -0.25 & 1.5/ 5.5  &\textrm{July 6, 2004} &  60 & 56.9 & 6.98 & -0.22 & 0.03 \\
          \textrm {\object{HD~40460}} & \textrm {K1III}  & 4741 & 2.00  & -0.50 & 1.5/ 5.0  &\textrm{Dec. 19, 2004} & 480 & 18.3  & 6.70 & -0.50 & 0.00 \\
          \textrm {\object{HD~81192}} & \textrm {G7III}  & 4705 & 2.50  & -0.62 & 1.5/ 5.5  &\textrm{April 6, 2004} & 512 & 12.7 & 6.73 & -0.47 & 0.15\\
          \textrm {\object{HD~117876}} & \textrm {G8III}  & 4782 & 2.25  & -0.50 & 1.5/ 5.0  &\textrm{April 5, 2004} & 480& 19.9 & 6.80 & -0.40 & 0.10 \\
          \textrm {\object{HD~139195}} & \textrm {K0p}  & 4856 & 2.82  & -0.04 & 1.5/ 5.5  &\textrm{May 17, 2004} & 88 & 13.0 & 7.02& -0.18 & -0.14 \\
          \textrm {\object{HD~161074}} & \textrm {K4III}  & 3980 & 1.73  & -0.27 & 1.5/ 6.0  &\textrm{June 22/23, 2005} & 960 & 60.9 & 7.13& -0.07 & 0.20\\
          \textrm {\object{HD~168723}} & \textrm {K0III-IV}  & 4859 & 3.13  & -0.19 & 1.5/ 4.7  &\textrm{May 17, 2004} & 20& 12.4 & 7.15& -0.05 & 0.14\\
            \textrm {\object{HD~184406}} & \textrm {K3III}  & 4520 & 2.41  & +0.01 & 1.5/ 5.0  &\textrm{Aug. 7, 2004} & 120 & 25.1 & 7.15& -0.05 & -0.06\\
          \textrm {\object{HD~188512}} & \textrm {G8IVvar}  & 5041 & 3.04  & -0.04 & 1.5/ 5.0  &\textrm{July 5, 2004} & 180& 9.1 & 6.98& -0.22 & -0.18\\
           \textrm {\object{HD~212943}} & \textrm {K0III}  & 4586 & 2.81  & -0.34 & 1.5/ 4.5  &\textrm{June 29, 2004} & 84 & 19.9 & 7.07& -0.13 & 0.21\\
          \textrm {\object{HD~214567}} & \textrm {G8II}  & 5038 & 2.50  & +0.03 & 1.5/ 5.5  &\textrm{June 30, 2004} & 288 & 19.7 & 7.15& -0.05 & -0.08 \\
          \textrm {\object{HD~219615}} & \textrm {G7III}  & 4830 & 2.57  & -0.42 & 1.5/ 6.0  &\textrm{June 29, 2004} & 64 & 19.6 &  7.02& -0.18 & 0.24 \\
          \textrm {\object{HD~220954}} & \textrm {K1III}  & 4664 & 2.37  & -0.10 & 1.5/ 5.5  &\textrm{June 29, 2004} & 48 & 30.0& 7.19& -0.01 & 0.09 \\
            \noalign{\smallskip}
            \hline
             \noalign{\smallskip}
         \textrm {HD~168723} & \textrm {K0III-IV}  & 4921 & 2.99  & -0.21 & 1.09/ 4.7  &\textrm{May 17, 2004} & 20& 12.4 & 7.07& -0.13 & 0.08\\
        \textrm {HD~188512} & \textrm {G8IVvar}  & 5110 & 3.60  & -0.17 & 0.92/ 5.0  &\textrm{July 5, 2004} & 180& 9.1 & 7.23& +0.03 & +0.20\\
            \hline
        \end{array}
     $$
   \end{table*}

For the analysis of the stellar spectra, 1D Marcs model atmospheres
(Gustafsson et al., in preparation) are computed for each star.
Sulphur abundances were derived from the equivalent widths, using
the Eqwi code, and the synthetic spectra were calculated with the
Bsyn code. Both codes use many subroutines from the Marcs code and
solve the radiative transfer at many frequencies in each spectral
line using the continuous opacity from the model atmosphere and the
line opacity calculated from line lists. The fundamental stellar
parameters of the observed stars, which are needed in the
calculation of the stellar atmospheres,
 are taken from \citet{elodie_II} and are also
given in Table \ref{spectrum}.
Moreover, a microturbulence of $1.5\,$\kms\ is used and
the usual linearly increased (for decreasing metallicity) $\alpha$ element enhancement for disk stars is assumed.
The Marcs models used are hydrostatic model photospheres in
spherical geometry (although not really necessary for the stars
investigated here) and are computed assuming LTE, chemical
equilibrium, homogeneity, and the conservation of the total flux
(radiative plus convective, the convective flux being computed using
the mixing length recipe). The  radiation field used in the model
generation is calculated with absorption from atoms and molecules by
opacity sampling at approximately 95\,000
wavelength points over the wavelength range $1300\,\mbox{\AA} $--$
20\,\mbox{$\mu$m}$. The atomic opacity sampling files are calculated
for metallicities of [Fe/H]$=0.0$, $-0.1$, $-0.25$, $-0.37$, $-0.5$,
and $-0.75$, and the one closest in metallicity is used.

Molecular lines are included, but only CN (line list from J\o
rgensen \& Larsson 1990, \nocite{jorg_CN} and Plez 1998, private
communications) is important for the most metal-rich stars with low
effective temperatures. TiO is, for example, unimportant even for
the coolest of the stars. Atomic line data are taken from the VALD
database (Piskunov et al. 1995)\nocite{VALD} with the exception of
the transition probability of the \sfor line, which is taken from
the NIST database version 3 (see, e.g., Martin et al.
2000)\nocite{NIST}. The wavelength and strength of the Fe\,{\sc i}
line and the position of the two Cr\,{\sc i} lines in the vicinity
of the \sfor line are matched to fit the solar lines (see Table
\ref{line_parameters}).
To take additional line broadening into account, the synthetic
spectra are convolved with a radial-tangential function (Gray 1992)
\nocite{gray} with a macroturbulent velocity given in Table
\ref{stars} as $\xi_\mathrm{macro}$. The synthetic spectra are used
to inspect
the location of the continuum, the line strengths and widths, and line blendings. For the stars we are investigating,
no weak atomic blends are found at the \sfor line.
The sulphur abundances are given in Table \ref{stars}.

  \begin{figure}
   \centering
  \includegraphics[width=8.5cm]{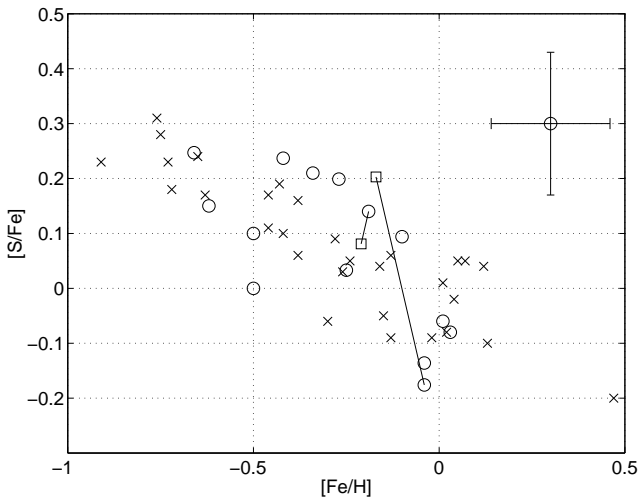}
      \caption{[S/Fe] vs. [Fe/H] plot for the stars investigated in this paper using the \sforbidden line (circles) and for
      the \citet{chen} stars (crosses). Typical error bars are indicated. Squares indicate new sulphur abundances derived
      from a different set of stellar parameters for $\eta$ Ser and $\beta$ Aql (see text).}
         \label{S_Fe_H}
   \end{figure}

In Fig. \ref{S_Fe_H} we have plotted our derived values of [S/Fe]
vs. [Fe/H], as well as the the relative sulphur abundances derived
for disk stars by \citet{chen}, who used optical, high-excitation
S\,{\sc i} lines ($\chi_\mathrm{exc}=7.9$ eV).

\section{Discussion}

We have chosen to use the stars in the library of \citet{elodie_II}
to keep to a set of stars with homogeneously determined stellar
parameters. The accuracies of the parameters given in
\citet{elodie_I} for the stars in this library are $\delta
T_\mathrm{eff}=86\,\mathrm K$, $\delta \log g=0.28 \,\mathrm {dex}$,
and $\delta \mathrm{[Fe/H]}=0.16 \,\mathrm {dex}$. In Table
\ref{error1} the uncertainties in the derived sulphur abundances due
to uncertainties in the fundamental stellar parameters are shown. We
also show the corresponding changes for the nearby  Cr\,{\sc i} and
Fe\,{\sc i} lines. Including the statistical measurement
uncertainties, as estimated from the continuum placement and the
effect of the SNR (both max 10\% in $W_\mathrm S$), a conservative
total uncertainty in the derived sulphur abundances
is estimated to be $0.13\,\mathrm{dex}$. 
The uncertainty in the metallicity is taken from \citet{elodie_I},
i.e., $\delta \mathrm{[Fe/H]}=0.16$ dex.

The strengths of weak lines depend on the line and continuum
opacities, $\chi_\mathrm{line}$ and $\chi_\mathrm{cont}$, such that
the equivalent width of a line
$W\propto\chi_\mathrm{line}/\chi_\mathrm{cont}$. Since sulphur is
mainly in the atomic state throughout the photospheres and since the
weak \sfor line originates in the ground state, the sulphur line
opacity (or the number of sulphur atoms in the ground state) is not
very temperature sensitive. The continuous opacity of the wavelength
region (H$^-_\mathrm{bf}$) is slightly $T$-sensitive, but we see
from Table \ref{error1} that the sulphur abundance derived is not
very $T$-sensitive. Nor is the sulphur line opacity very sensitive
to the electron pressure, although the continuous opacity is. This
explains the sensitivity of the abundance derived from the observed
equivalent width to the metallicity and especially to the surface
gravity, as seen in Table \ref{error1}. The Cr and Fe lines behave
differently since they are from excited states, and (like H$^-$)
represent a minority species in the photosphere. Therefore, the
derived Cr and Fe abundances are more temperature sensitive, but are
not as sensitive to the electron pressure. It can also be noted that
the \sfor line is not saturated, and therefore not sensitive to
$\xi_\mathrm{micro}$.

In the discussion on errors we have ignored systematic errors due to
the treatment of convection in the atmospheres. A more proper way of
treating convection is through ab initio 3D hydrodynamic models as
described in, e.g., \citet{asplund:3D1}. For the [O\,{\sc i}] lines,
\citet{nissen:oi} show that for disk metallicities the 3D effects
are less than 0.1 dex (decreasing with increasing metallicity) for
their coolest dwarf (5800 K). The main effect is the change in the
continuous opacity and not a change in the line opacity due to
cooler upper layers. Sulphur is expected to behave in a similar
fashion. The effect is unknown for giants and subgiants, and we did
not include it in the present analysis.

In his careful analysis of nearby disk and halo stars,
\citet{fuhrmann:04} has two stars in common with our star list,
namely $\eta$ Ser (HD168723) and $\beta$ Aql (HD188512). The stellar
parameters derived by him, and the alternative sulphur abundances
derived by us using his stellar parameters, are also given in Table
\ref{stars}. As we saw in our uncertainty analysis, the most
sensitive fundamental parameter of the stellar models influencing
the abundance derived from the \sfor line is the surface gravity. In
particular, the alternative surface gravity of HD188512 ($\log g =
3.04 \rightarrow 3.60$) is the major cause of the large change in
the sulphur abundance of that star ($\log \varepsilon(\mathrm
S)=6.98 \rightarrow 7.23$). Since the metallicity of HD188512 is
also different ([Fe/H]$-0.04 \rightarrow -0.17$), the position of
the star in Fig. \ref{S_Fe_H} changes quite dramatically. This again
gives an indication of the uncertainties due to the stellar
parameters only.

From Fig. \ref{S_Fe_H} we see that the agreement with the abundance
determination of \citet{chen} is good given the level of
uncertainties. \citet{chen} showed that their S abundances
correlated strongly with their Si abundances, the latter being a
well studied $\alpha$ element, suggesting that the nucleosynthesis
of sulphur is similar to that of the other $\alpha$ elements. We can
thus corroborate that sulphur shows an evolution typical of an
$\alpha$ element, and we have shown that the \sforbidden line is
indeed a good diagnostic tool for the sulphur abundance in disk
stars. The virtues of the line are that it is not very sensitive to
$T_\mathrm{eff}$ and that it should not be subject to non-LTE
effects. It is apparent that an uncertainty in the surface gravity
affects the abundance derived from this line. It should, however, be
noted that the [S/Fe] ratio will be much less dependant on gravity
if the iron abundance is determined from Fe\,{\sc ii} lines that
react similarly to a change in electron pressure as the \sfor line.
The question whether the \sforbidden line can also be used for stars
of the galactic halo will be addressed in a forthcoming paper.

\begin{table}
      \caption[]{Atomic line parameters. After the excitation energy, $\chi_\mathrm{exc}$, of the lower level of the transition,
      which is given in eV,
      the transition probability of the line as the logarithm of the product of the statistical weight and the
      oscillator strength is given.
      }
         \label{line_parameters}
     $$
         \begin{array}{lcccc}
            \hline
            \hline
            \noalign{\smallskip}
     \mathrm{Element} &   \mathrm{Wavenumber} & \mathrm{Wavelength}     & \textrm{$\chi_\mathrm{exc}$} & \textrm{$\log\,gf$}   \\
            \noalign{\smallskip}
            & & \mathrm{(in\,\,vacuum)}\\
            \noalign{\smallskip}
        & \textrm{[cm$^{-1}$]} & \textrm{[\AA ]} & \textrm{[eV]} &  \\
           \noalign{\smallskip}
              \hline
           \noalign{\smallskip}
\textrm {Cr \sc{i}}& 9242.264 & 10819.860  &   3.013  &  -1.957 \\
\textrm {Fe \sc{i}} & 9241.082 & 10821.244  &   3.960   &   -2.178 \\
\textrm {[S \sc{i}]} & 9238.609 & 10824.140   &  0.000  &     -8.617 \\ 
\textrm {Cr \sc{i}} & 9238.200 & 10824.619 & 3.013  &  -1.678  \\
            \noalign{\smallskip}
            \hline
           \noalign{\smallskip}
         \end{array}
     $$
   \end{table}

\begin{table}
  \caption{A representative example of the effects on logarithmic abundances derived for the S, Cr, and Fe lines in the
  observed wavelength region when the fundamental
  parameters of the model atmosphere of HD117876 are changed.}
  \label{error1}
$$
  \begin{tabular}{l c c c c   }
           \hline
            \hline
   \noalign{\smallskip}
   Uncertainty & $\mathrm { \Delta \log\varepsilon(S) }$ & $\mathrm { \Delta \log\varepsilon(Cr) }$ & $\mathrm { \Delta \log\varepsilon(Fe)}$\\
  \noalign{\smallskip}
  \hline
  \noalign{\smallskip}
  $\delta T_\mathrm{eff}=-100\, \mathrm { K }$ &  $-0.02$ & $-0.09$& $-0.05$  \\
 \noalign{\smallskip}
  $ \delta \log g =-0.25  $ & $-0.11$ & $\pm0.00$  & $-0.02$ \\
  \noalign{\smallskip}
 $\mathrm { \delta [Fe/H]=-0.15  }$ &  $-0.06$ & $\pm0.00$& $-0.01$ \\
  \noalign{\smallskip}
 $\mathrm { \delta \xi_{\rm micro}=-0.5  }$ &  $\pm0.00$ & $´+0.07$ & $+0.06$\\
  \hline
  \end{tabular}
$$
\end{table}

\begin{acknowledgements}
I am grateful to the referee, P. Bonifacio, for insightful comments
and suggestions, K. Eriksson and B. Gustafsson for fruitful
discussions on model atmospheres, and D. L. Lambert for help in
planning the observations. I thank the Gemini Staff, particularly K.
H. Hinkle and V. V. Smith, for their support during the
observations. This work was supported by the Swedish Research
Council, \emph{VR}.
\end{acknowledgements}


\begin{thebibliography}{}

\bibitem[\protect\astroncite{{Asplund} et~al.}{2004}]{asplund_oi1}
{Asplund}, M., {Grevesse}, N., {Sauval}, A.~J., {Allende Prieto},
C., \& {Kiselman}, D. 2004,
\newblock {A\&A} {417}, 751

\bibitem[\protect\astroncite{{Asplund} et~al.}{2000}]{asplund:3D1}
{Asplund}, M., {Nordlund}, {\AA}., {Trampedach}, R., {Allende
Prieto}, C., \&
  {Stein}, R.~F. 2000
\newblock {A\&A} {359}, 729

\bibitem[\protect\astroncite{{Caffau} et~al.}{2005}]{caffau}
{Caffau}, E., {Bonifacio}, P., {Faraggiana}, R., {Fran{\c c}ois}, P.,
  {Gratton}, R.~G., \& {Barbieri}, M. 2005,
\newblock {A\&A} {441}, 533

\bibitem[\protect\astroncite{{Chen} et~al.}{2002}]{chen}
{Chen}, Y.~Q., {Nissen}, P.~E., {Zhao}, G., \& {Asplund}, M. 2002,
\newblock {A\&A} {390}, 225

\bibitem[\protect\astroncite{{Clegg} et~al.}{1981}]{clegg}
{Clegg}, R.~E.~S., {Tomkin}, J., \& {Lambert}, D.~L. 1981,
\newblock {ApJ} {250}, 262

\bibitem[\protect\astroncite{{Francois}}{1987}]{francois:87}
{Francois}, P. 1987,
\newblock {A\&A} {176}, 294

\bibitem[\protect\astroncite{{Fuhrmann}}{2004}]{fuhrmann:04}
{Fuhrmann}, K. 2004,
\newblock {Astronomische Nachrichten} {325}, 3

\bibitem[\protect\astroncite{{Garc{\'{\i}}a P{\'e}rez} et~al.}{2006}]{perez_oi}
{Garc{\'{\i}}a P{\'e}rez}, A.~E., {Asplund}, M., {Primas}, F., {Nissen}, P.~E.,
  \& {Gustafsson}, B. 2006,
\newblock {A\&A} {451}, 621

\bibitem[\protect\astroncite{{Gray}}{1992}]{gray}
{Gray}, D.~F. 1992,
\newblock {{The observation and analysis of stellar photospheres}}
\newblock (New York: Cambridge Univ. Press)

\bibitem[\protect\astroncite{{Hinkle} et~al.}{1998}]{phoenix}
{Hinkle}, K.~H., {Cuberly}, R.~W., {Gaughan}, N.~A., et~al. 1998,
\newblock {SPIE} {3354}, 810

\bibitem[\protect\astroncite{{Israelian} and {Rebolo}}{2001}]{israel}
{Israelian}, G. \& {Rebolo}, R. 2001,
\newblock {ApJ} {557}, L43

\bibitem[\protect\astroncite{{Jorgensen} and {Larsson}}{1990}]{jorg_CN}
{Jorgensen}, U.~G. \& {Larsson}, M. 1990,
\newblock {A\&A} {238}, 424

\bibitem[\protect\astroncite{{Katz} et~al.}{1998}]{elodie_I}
{Katz}, D., {Soubiran}, C., {Cayrel}, R., {Adda}, M., \& {Cautain},
R. 1998,
\newblock {A\&A} {338}, 151

\bibitem[\protect\astroncite{{Korn} and {Ryde}}{2005}]{korn_ryde:05}
{Korn}, A.~J. \& {Ryde}, N. 2005,
\newblock {A\&A} {443}, 1029

\bibitem[\protect\astroncite{{Martin} et~al.}{2000}]{NIST}
{Martin}, W.~C., {Fuhr}, J., {Kelleher}, D., et al. 2000,
\newblock {Atomic and Molecular Data for Astrophysics: New Developments, Case
  Studies and Future Needs, 24th IAU meeting, JD1}

\bibitem[\protect\astroncite{{Nissen} et~al.}{2004}]{PEN_halo}
{Nissen}, P.~E., {Chen}, Y.~Q., {Asplund}, M., \& {Pettini}, M.
2004,
\newblock {A\&A} {415}, 993

\bibitem[\protect\astroncite{{Nissen} et~al.}{2002}]{nissen:oi}
{Nissen}, P.~E., {Primas}, F., {Asplund}, M.,  \& {Lambert}, D.~L.
2002,
\newblock {A\&A} {390}, 235

\bibitem[\protect\astroncite{{Piskunov} et~al.}{1995}]{VALD}
{Piskunov}, N.~E., {Kupka}, F., {Ryabchikova}, T.~A., {Weiss},
W.~W., \&
  {Jeffery}, C.~S. 1995,
\newblock {A\&AS} {112}, 525

\bibitem[\protect\astroncite{{Ryde} et~al.}{2005}]{ryde_munchen_review}
{Ryde}, N., {Gustafsson}, B., {Eriksson}, K., \& {Wahlin}, R. 2005,
\newblock in {HR IR Spectroscopy in Astronomy}, ed. H.~U. {K{\"a}ufl}, R. {Siebenmorgen},
\& A.~F.~M. {Moorwood} (Berlin/Heidelberg: Springer), 365

\bibitem[\protect\astroncite{{Ryde} and {Lambert}}{2004}]{ryde:04_S}
{Ryde}, N. \& {Lambert}, D.~L. 2004,
\newblock {A\&A} {415}, 559

\bibitem[\protect\astroncite{{Ryde} and {Lambert}}{2005}]{ryde:05_S_poster}
{Ryde}, N. \& {Lambert}, D.~L. 2005,
\newblock in {Cosmic Abundances as Records of Stellar Evolution and
Nucleosynthesis}, ed. T.~G. {Barnes} \& F.~N. {Bash} (ASP Conf. Ser.
336), 355

\bibitem[\protect\astroncite{{Soubiran} et~al.}{1998}]{elodie_II}
{Soubiran}, C., {Katz}, D., \& {Cayrel}, R. 1998,
\newblock {A\&AS} {133}, 221

\bibitem[\protect\astroncite{{Takada-Hidai} et~al.}{2002}]{takeda}
{Takada-Hidai}, M., {Takeda}, Y., {Sato}, S., et~al. 2002,
\newblock {ApJ} {573}, 614

\bibitem[\protect\astroncite{{Tody}}{1993}]{IRAF}
{Tody}, D. 1993,
\newblock in Astronomical Data Analysis Software and Systems II, ed. R.~J. {Hanisch},
R.~J.~V. {Brissenden}, \& J. {Barnes} (ASP Conf. Ser. 52), 173

\end{thebibliography}
\bibliographystyle{aabib99}



\end{document}